\documentclass[prd,superscriptaddress,amsfonts,amssymb,amsmath,showpacs,twocolumn]{revtex4-1}

\usepackage{amsmath}
\usepackage{epsfig}
\usepackage{graphicx}
\usepackage[caption=false,font=footnotesize,labelformat=simple]{subfig}
\usepackage{xcolor}
\usepackage{color}
\usepackage[colorinlistoftodos]{todonotes}
\usepackage{xparse}
\usepackage[colorlinks=true,
            linkcolor=black,
            urlcolor=blue,
            citecolor=blue]{hyperref}

\ExplSyntaxOn
\NewDocumentCommand{\mref}{m}{\quinn_mref:n {#1}}
\seq_new:N \l_quinn_mref_seq
\cs_new:Npn \quinn_mref:n #1
 {
  \seq_set_split:Nnn \l_quinn_mref_seq { , } { #1 }
  \seq_pop_right:NN \l_quinn_mref_seq \l_tmpa_tl
  ( 
  \seq_map_inline:Nn \l_quinn_mref_seq
    { \ref{##1},\nobreakspace } 
  \exp_args:NV \ref \l_tmpa_tl 
  ) 
 }
\ExplSyntaxOff
\begin{document}

\title{Gravitational lensing by wormholes supported by electromagnetic,
scalar, and quantum effects}

\author{Kimet Jusufi}
\email{kimet.jusufi@unite.edu.mk}
\affiliation{Physics Department, State University of Tetovo, Ilinden Street nn, 1200,
Tetovo, MACEDONIA}
\affiliation{Institute of Physics, Faculty of Natural Sciences and Mathematics, Ss. Cyril and Methodius University, Arhimedova 3, 1000 Skopje, MACEDONIA}

\author{Ali \"{O}vg\"{u}n}
\email{ali.ovgun@pucv.cl}
\homepage{http://www.aovgun.com}
\affiliation{Instituto de F\'{\i}sica, Pontificia Universidad Cat\'olica de
Valpara\'{\i}so, Casilla 4950, Valpara\'{\i}so, CHILE}

\affiliation{Department of Physics and Astronomy, University of Waterloo, Waterloo, Ontario, N2L 3G1, CANADA}
\affiliation{Perimeter Institute for Theoretical Physics, Waterloo, Ontario, N2L 2Y5, CANADA}
\affiliation{Physics Department, Arts and Sciences Faculty, Eastern Mediterranean University, Famagusta, North Cyprus via Mersin 10, TURKEY}

\author{ Ayan Banerjee}
\email{ayan\_7575@yahoo.co.in}
\affiliation{Department of Mathematics, Jadavpur University, Kolkata 700032, West Bengal, INDIA}
\affiliation{Astrophysics and Cosmology Research Unit, University of KwaZulu Natal, Private Bag X54001, Durban 4000,
SOUTH AFRICA}

\author{\.{I}zzet Sakall\i{}}
\email{izzet.sakalli@emu.edu.tr}
\affiliation{Physics Department, Arts and Sciences Faculty, Eastern Mediterranean University, Famagusta, North Cyprus via Mersin 10, TURKEY}
\date{\today }

\begin{abstract}
Wormholes are one of the most interesting topological features in spacetime,
offering a rat run between two vastly separated regions of the universe. In
this paper, we study the deflection angle of light by wormholes, which are
supported by electric charge, magnetic charge, and scalar fields in the weak
field limit approximation. To this end, we apply new geometric methods --
the Gauss-Bonnet theorem and the optical geometry -- to compute the
deflection angles. We also verify our findings by using the well-known
geodesics method. There exists a similarity between the charge and the
quantum corrections on a black hole solution, which has been recently
discussed in the context of the relativistic Bohmian quantum mechanics. By
replacing classical geodesics with Bohmian trajectories, we introduce a new
wormhole solution, whose having matter sources and anisotropic pressure
supported by Bohmian quantum effects. The problem of fulfillment of the
energy conditions of the Morris-Thorne traversable wormhole is also
discussed.
\end{abstract}

\maketitle
\tableofcontents

\section{Introduction}

During the last few decades, wormholes have become one of the most popular
and intensively studied topics. They are tunnel-like connections made out of
spacetime, offering a shorter distance between two vastly separated regions
of spacetime (of the different universes or widely separated areas of our
own Universe). In 1988, the seminal works on traversable wormholes, as
hypothetical shortcuts in spacetime, was proposed by Morris and Thorne \cite%
{Morris}. The wormhole solution culminated in the Visser's book with title "\textit{Lorentzian
Wormholes: From Einstein to Hawking"} \cite{Visser1}, which reviews the subject up to the
year 1995, as well as new notions are presented and hinted at. However,
wormhole physics dates back to 1935, when Einstein and Rosen \cite{Einstein}
constructed an elementary particle model represented by a bridge linking two
identical sheets: Einstein-Rosen bridge (ERB). But, ERB was an unsuccessful
particle model. In 1962, Fuller and Wheeler \cite{FullerWheeler} also showed
ERB is unstable if it connects two parts of the same universe. At this
juncture, it is worth noting that terminology of ``wormhole'' was first
coined by Wheeler \cite{whel}. In fact, Wheeler considered wormholes as quantum foams
(at the Planck scale), which link separate regions of spacetime \cite%
{Wheeler,Wheeler1}. In fact, ERB is a non-traversable wormhole, even by a
photon, since the throat pinches off in a finite time. These entities were
further explored by Ellis in 1973 \cite{Ellis,Ellis1}, by producing a
non-singular wormhole solutions using a wrong-sign scalar field. Later,
Thorne together with his students came up with a time-machine model using
wormholes \cite{Yurtsever}.

Morris and Thorne studied the ``traversable wormholes'' for human travel 
\cite{Morris}. They modeled the wormholes as the objects possessing two
mouths and a throat. It was understood that for fabricating such a wormhole
solutions, two spacetimes that do not posses any event horizon or physical
singularities should be invoked. But the wormhole throat is threaded by
exotic matter, which violates the null energy condition (NEC) \cite{Visser1}. The existence of the so-called exotic matter sounds to
be unusual in general relativity (GR), but such a matter appears in quantum
field theory as well as on scalar-tensor theories. 

A first example of rotating axially symmetric solution describing a traversable wormhole was
given by Teo \cite{Teo}. However, Teo only showed the generic form (together
with some physical properties) of such a rotating wormhole without
representing the solution to the associated Einstein-field equations. In a
similar way Khatsymovsky \cite{Khatsymovsky} discussed some general features
of a slowly rotating wormhole. Meanwhile, traversable Lorentzian wormhole in
the presence of a generic cosmological constant $\Lambda$ was studied by
Lemos et al. \cite{Lemos}, and further wormhole in anti-de Sitter background
was found by \cite{Lemos1}. On the other hand, wormhole solutions were
studied in the framework of modified and alternative theories of gravity
such as the braneworld scenario \cite{Lobo5,Chakraborty}, Einstein Gauss-Bonnet Gravity \cite%
{Mehdizadeh,Rani,Ric1,Halali1,Halali2,gol,Cl1,Cl2,Cl3,Cl4,Gibwh,HW,Son,Ki}, $f(R)$ gravity \cite{Lobo,Sharif,Bahamonde}%
, $f(R,\phi)$ gravity \cite{Zubair}, $f(R,T)$ gravity \cite%
{Moraes,Yousaf,Moraes1,Zubair1}, non-commutative gravity \cite%
{Kuhfittig,Sharif1,Rahaman,Bhar}.

A decade ago, new global topology method was suggested by Gibbons and Werner 
\cite{GibbonsWerner1,GibbonsWerner2} to calculate the deflection angle for
static black holes, in the weak limits, by using the Gauss-Bonnet theorem
(GBT). Then, Werner \cite{Werner3} also extended the latter result to the
stationary black holes. Afterwards, Gibbons and Werner method (GWM) has
become popular and is still used in many studies \cite{K1,K2,K3,S1,K4,K5,K6,Tono,K7,K8,K9,K10}. Different than the
ordinary methods, in GWM the deflection angle is computed by taking the
integral over a domain $A_{\infty }$ outside the light ray: 
\begin{equation}
\hat{\alpha}=-\int \int_{\mathcal{A}_{\infty }}\mathcal{K}\mathrm{d}\sigma.  \label{1n}
\end{equation}%
Note that $\mathcal{K}$ stands for the Gaussian optical curvature, $\mathrm{d}\sigma $ is the elementary surface area of the optical geometry, and $\hat{%
\alpha}$ is the deflection angle.

Despite the success of GR, quantum gravity has become dramatically urgent
necessity in order to resolve several problems in cosmos like the missing
matter problem or singularities of spacetime. In an recent approach, Das 
\cite{DasPRD} showed that by replacing the classical geodesics with quantal
trajectories arising from Bohmian mechanics, one obtains the quantum
corrections to the Raychaudhuri equation, which is the so-called quantum
Raychaudhuri equation (QRE). QRE has recently been employed for resolving
the singularities in cosmology \cite{Ali} and in black holes \cite{Ali1}. In
particular, in the study of \cite{Ali1} modified Schwarzschild metric has
been derived and, in sequel, the singularity structure and thermodynamical
features of the obtained black hole have been studied in detail. Remarkably,
it was shown that quantum corrections change the picture of Hawking
radiation when the size of BH approaches the Planck scale.

Therefore our preliminary aim of this study is to calculate the deflection angle of wormhole with electric charge, magnetic charge and next, scalar fields as well as the quantum effects in the weak field limit approximation using the GWM to do so we use the GBT. The main importance of the paper is to find the difference between the deflection angles which may shed light from an experimental point of view in the near future.

This paper has the following organization. In Sec. I, we consider a charged
wormhole and study the deflection angle of light from it. Section II is
devoted to the deflection angle computation for a wormhole spacetime
supported by nonlinear electromagnetic theory. In Sec. III, we perform the
same analysis in the case of a scalar field wormhole. In Sec. IV, we
introduce a new wormhole solution based on the analogy of charge and quantum
effects in the framework of the relativistic Bohmian quantum mechanics.
Furthermore, we investigate the deflection angle as well as the energy
conditions. We draw our conclusions in Sec. V.

\section{Deflection of light by Wormhole with electric charge}

Recently, Kim and Lee \cite{sung} have found an exact solution for a
charged Lorentzian wormhole, whose spacetime is described by the
following metric 
\begin{equation}
\mathrm{d}s^{2}=-\left( 1+\frac{Q^{2}}{r^{2}}\right) \mathrm{d}t^{2}+\left( 1-\frac{b(r)}{r}+%
\frac{Q^{2}}{r^{2}}\right) ^{-1}\mathrm{d}r^{2}+r^2\mathrm{d}\Omega _{2}^{2}.  \label{2chargedm}
\end{equation}

Throughout the paper, we shall consider the special case when $
b(r)=b_{0}^{2}/r$. Thus, the charged wormhole metric \eqref{2chargedm}
recasts \cite{sung}

\begin{equation}
\mathrm{d}s^{2}=-\left( 1+\frac{Q^{2}}{r^{2}}\right) \mathrm{d}t^{2}+\left( 1-\frac{b_{0}^{2}}{%
r^{2}}+\frac{Q^{2}}{r^{2}}\right) ^{-1}\mathrm{d}r^{2}+r^2\mathrm{d}\Omega _{2}^{2}.  \label{3}
\end{equation}

To find the charged wormhole optical metric, we simply let $\mathrm{d}%
s^{2}=0 $. Furthermore, it is quite convenient to simplify the problem by considering the deflection of light in the equatorial plane; hence we set $\theta =\pi /2$. The optical metric reads 
\begin{equation}
\mathrm{d}t^{2}=\frac{\mathrm{d}r^{2}}{\left( 1+\frac{Q^{2}}{r^{2}}\right)
\left( 1-\frac{b_{0}^{2}}{r^{2}}+\frac{Q^{2}}{r^{2}}\right) }+\frac{r^{2}}{%
\left( 1+\frac{Q^{2}}{r^{2}}\right) }\mathrm{d}\varphi ^{2}.  \label{4}
\end{equation}

At this point, one may introduce a new radial coordinate $u$, and a
new function $\zeta(u)$ as follows 
\begin{equation}
\mathrm{d}u=\frac{\mathrm{d}r}{\sqrt{\left( 1+\frac{Q^{2}}{r^{2}}%
\right) \left( 1-\frac{b_{0}^{2}}{r^{2}}+\frac{Q^{2}}{r^{2}}\right) }}%
,\,\,\,\zeta(u)=\frac{r}{\sqrt{1+\frac{Q^{2}}{r^{2}}}}.  \label{5}
\end{equation}

Thus, our metric can be rewritten as 
\begin{equation}
\mathrm{d}t^{2}=h_{ab}\,\mathrm{d}\lambda^{a}\mathrm{d}\lambda^{b}=\mathrm{d}u^2+\zeta^2(u)\mathrm{d}\varphi
^{2}.  \label{6}
\end{equation}

Note that $(a,b=r,\varphi )$ and $h=\det h_{ab}$. The Gaussian optical curvature $\mathcal{K}$, is defined by: 
\begin{equation}
\mathcal{K}=-\frac{1}{\zeta(u)}\left[ \frac{\mathrm{d}r}{\mathrm{d}u}%
\frac{\mathrm{d}}{\mathrm{d}r}\left( \frac{\mathrm{d}r}{\mathrm{d}u}%
\right) \frac{\mathrm{d}\zeta}{\mathrm{d}r}+\left( \frac{\mathrm{d}r}{\mathrm{d}%
u}\right) ^{2}\frac{\mathrm{d}^{2}\zeta}{\mathrm{d}r^{2}}\right] .
\label{7curvature}
\end{equation}

Applying Eq. \eqref{7curvature} to our metric \eqref{3}, we find the charged wormhole
Gaussian optical curvature, which is given by 
\begin{equation}
\mathcal{K}={\frac{2{Q}^{6}+5{Q}^{4}{r}^{2}-2{Q}^{4}{b_0^2+3{Q}^{2}{r}%
^{4}-5{Q}^{2}{r}^{2}b_0^2-{r}^{4}b_0^2}}{{r}%
^{6}\left( {Q}^{2}+{r}^{2}\right) }}.  \label{8}
\end{equation}

We can approximate the above result in leading orders. Thus, one gets 
\begin{equation}
\mathcal{K}={\frac{3\,{Q}^{2}-b_0^2}{{r}^{4}}}-{\frac{4\,b_0^2\,{Q}^{2}}{{r}^{6}}}+\mathcal{O}(Q^{4},b_{0}^{4}).  \label{9}
\end{equation}

This result is going to be used to evaluate the deflection angle. The key
point behind the GWM is the GBT, which involves the wormhole optical
geometry in our case. More precisely, we can choose a non-singular domain
outside the light ray, let us say $\mathcal{A}_{R}$, with boundary $\partial 
\mathcal{A}_{R}=\gamma _{h}\cup C_{R}$, from where the GBT can be
expressed as 
\begin{equation}
\iint\limits_{\mathcal{A}_{R}}\mathcal{K}\,\mathrm{d}\sigma+\oint\limits_{\partial \mathcal{%
A}_{R}}\kappa \,\mathrm{d}t+\sum_{i}\theta _{i}=2\pi \chi (\mathcal{A}_{R}).
\label{10}
\end{equation}

Note that $\kappa $ gives the geodesic curvature, $\mathcal{K}$ stands for the
Gaussian optical curvature, while $\theta _{i}$ is the exterior angle at the 
$i^{th}$ vertex. We can choose a non-singular domain outside of the light
ray with the Euler characteristic number $\chi (\mathcal{A}_{R})=1$. The
geodesic curvature is defined as 
\begin{equation}
\kappa =h\,\left( \nabla _{\dot{\gamma}}\dot{\gamma},\ddot{\gamma}%
\right),  \label{11}
\end{equation}%
where the unit speed condition $h(\dot{\gamma},\dot{\gamma})=1$,
holds. If we let $R\rightarrow \infty $, our two jump angles ($\theta _{%
\mathcal{O}}$, $\theta _{\mathcal{S}}$) become $\pi /2,$ or in other words,
the sum of jump angles to the source $\mathcal{S}$, and observer $\mathcal{O}
$, satisfies $\theta _{\mathit{O}}+\theta _{\mathit{S}}\rightarrow \pi $ 
\cite{GibbonsWerner1}. Hence, we can rewrite the GBT as 
\begin{equation}
\iint\limits_{\mathcal{A}_{R}}\mathcal{K}\,\mathrm{d}\sigma+\oint\limits_{C_{R}}\kappa \,%
\mathrm{d}t\overset{{R\rightarrow \infty }}{=}\iint\limits_{\mathcal{D}%
_{\infty }}\mathcal{K}\,\mathrm{d}\sigma+\int\limits_{0}^{\pi +\hat{\alpha}}\mathrm{d}%
\varphi =\pi .  \label{12}
\end{equation}

Let us now compute the geodesic curvature $\kappa $. To this end, we first
point out that $\kappa (\gamma _{h})=0$, since $\gamma _{h}$
is a geodesic. Thus, we end up with 
\begin{equation}
\kappa (C_{R})=|\nabla _{\dot{C}_{R}}\dot{C}_{R}|,  \label{13}
\end{equation}%
in which one can choose $C_{R}:=r(\varphi )=R=\text{const}$. The radial part
is evaluated as follows 
\begin{equation}
\left( \nabla _{\dot{C}_{R}}\dot{C}_{R}\right) ^{r}=\dot{C}_{R}^{\varphi
}\,\left( \partial _{\varphi }\dot{C}_{R}^{r}\right) +\Gamma%
_{\varphi \varphi }^{r(op)}\left( \dot{C}_{R}^{\varphi }\right) ^{2}.  \label{14}
\end{equation}

In the last equation $\Gamma%
_{\varphi \varphi }^{r(op)}$ gives the Christoffel symbol which is associated to the optical geometry. From the above equation, it is obvious that the first term vanishes, while
the second term is calculated using Eq. \eqref{13} and the unit speed
condition. As $R\rightarrow \infty ,$ the geodesic curvature and $\mathrm{d}%
t $ approximates to%
\begin{eqnarray}
\lim_{R\rightarrow \infty }\kappa (C_{R}) &=&\lim_{R\rightarrow \infty
}\left\vert \nabla _{\dot{C}_{R}}\dot{C}_{R}\right\vert ,  \notag \\
&\rightarrow &\frac{1}{R}.  \label{15n}
\end{eqnarray}

and 
\begin{eqnarray}
\lim_{R\rightarrow \infty }\mathrm{d}t &=&\lim_{R\rightarrow \infty }\left( 
\frac{R}{\sqrt{1+\frac{Q^{2}}{R^{2}}}}\right) \mathrm{d}\varphi  \notag \\
&\rightarrow &R\,\mathrm{d}\varphi .  \label{16n}
\end{eqnarray}

If we combine Eqs. \eqref{15n} and \eqref{16n}, we find $\kappa (C_{R})%
\mathrm{d}t=\mathrm{d}\varphi $. For getting the deflection angle, it is
convenient to approximate the boundary curve of $\mathcal{A}_{\infty }$ to a
notional undeflected ray, which is the line of $r(\varphi )=b/\sin \varphi $%
. Then, the deflection angle \eqref{1n} reads 
\begin{equation}
\hat{\alpha}=-\int\limits_{0}^{\pi }\int\limits_{\frac{b}{\sin \varphi }%
}^{\infty }\mathcal{K}\mathrm{d}\sigma.  \label{17n}
\end{equation}

If we substitute Eq. \eqref{9} into \eqref{17n}, we find out the following
integral

\begin{equation}
\hat{\alpha}=-\int\limits_{0}^{\pi }\int\limits_{\frac{b}{\sin \varphi }%
}^{\infty }\left( {\frac{3\,{Q}^{2}-b_0^2}{{r}^{4}}}-{\frac{4\,b_0^2\,{Q}^{2}}{{r}^{6}}}\right) \sqrt{\det h_{ab}}\,%
\mathrm{d}r\mathrm{d}\varphi .  \label{18n}
\end{equation}

It is worth noting that using the relation: $\mathrm{d}u \approx 
\mathrm{d}r$, which is valid in the limit $R\rightarrow \infty $, one can
easily solve \eqref{18n} in the leading order terms. Thus, we have 
\begin{equation}
\hat{\alpha}\simeq \frac{\pi b_{0}^{2}}{4\,b^{2}}-\frac{3\pi Q^{2}}{4\,b^{2}}%
+\mathcal{O}(Q^2,b_0^2)
\end{equation}

Thus, the total deflection angle consists by two leading terms, the first terms is a direct result of the  geometry of the spacetime, while the second term a consequence of the electric  field.

\subsection{Geodesics of Wormhole with Electric Charge}

Lagrangian of metric (2) is expressed as follows 
\begin{eqnarray}\label{20n}\notag
2\,\mathcal{L}&=&-\left(1+\frac{Q^2}{r(s)^2}\right)\dot{t}^2(s)+\frac{\dot{r}(s)^2}{1-\frac{b_0^2}{r(s)^2}+\frac{Q^2}{r(s)^2}}\\
&+&r(s)^2\dot{\theta}^2(s)+r(s)^2\sin^2\theta \dot{\varphi}^2(s).
\end{eqnarray}

As it is well known, the geodesic equation has to be supplemented by the
normalization (or the so-called metric) condition $2\mathcal{L}=\epsilon $
where for massive particles $\epsilon =1$ and for photon $\epsilon =0$.
Here, we consider the deflection of planar photons i.e., $\theta =\pi /2$.
After using the spacetime symmetries, one obtains two constants of motion $l$
and $\gamma $, which are given by \cite{geodesicsBoyer}: 
\begin{eqnarray}
p_{\varphi } &=&\frac{\partial \mathcal{L}}{\partial \dot{\varphi}}=2r^{2}(s)%
\dot{\varphi}(s)=\mathcal{G},  \label{21n} \\
p_{t} &=&\frac{\partial \mathcal{L}}{\partial \dot{t}}=-2\left( 1+\frac{Q^{2}%
}{r(s)^{2}}\right) \dot{t}(s)=-\mathcal{E}.  \label{22n}
\end{eqnarray}

Let us now introduce a new variable $u(\varphi )$, which is related to our
old radial coordinate as $r=1/u(\varphi )$. Thus, we obtain the following
identity: 
\begin{equation}
\frac{\dot{r}}{\dot{\varphi}}=\frac{\mathrm{d}r}{\mathrm{d}\varphi }=-\frac{1%
}{u^{2}}\frac{\mathrm{d}u}{\mathrm{d}\varphi }.  \label{23n}
\end{equation}

Without loss of generality, we can normalize the affine parameter along the
light ($\epsilon =0$) rays by taking $\mathcal{E}=1$ \cite{geodesicsBoyer}
and approximate the distance of closest approach with the impact parameter
i.e., $u_{max}=1/r_{min}=1/b$, since we shall consider only leading order
terms \cite{Iorio}. In this case, one can choose the second constant of
motion: $\mathcal{G}=b$. 

Finally using the metric condition ($\epsilon =0$) with Eqs. (20-23) in the associated Lagrange equation of $u(\varphi )$, we obtain the following
equation: 
\begin{eqnarray}\notag
&&\left(\frac{\mathrm{d}u}{\mathrm{d}\varphi}\right)^2 \frac{1}{(Q^2 u^2-b_0^2 u^2+1)u^4}-\frac{Q^2}{u^6 u^2 (Q^2+\frac{1}{u^2})^2}\\
&-&\frac{1}{u^8 b^2 (Q^2+\frac{1}{u^2})^2}+\frac{1}{u^2}=0.
\end{eqnarray}

From here, it follows that 
\begin{eqnarray}
\frac{\mathrm{d}\varphi}{\mathrm{d}u}=\pm \frac{b \sqrt{Q^2 u^2 +1}}{\sqrt{b^2\Phi(Q,u,b_0)+Q^2u^2-b_0^2 u^2 +1}},
\end{eqnarray}
where 
\begin{equation}\notag
\Phi(Q,u,b_0)=-Q^4u^6+Q^2 b_0^2 u^6-2Q^2 u^4+b_0^2 u^4-u^2.
\end{equation}

It is well known that the solution of the differential equation (45) is
given by the following relation \cite{geodesicsBoyer,weinberg}

\begin{equation}
\Delta \varphi =\pi +\hat{\alpha},  \label{26n}
\end{equation}%
where $\hat{\alpha}$ is the deflection angle to be derived. Following the
same arguments given in Ref. \cite{weinberg}, the deflection angle can be
calculated as 
\begin{equation}
\hat{\alpha}=2|\varphi _{u={1/b}}-\varphi _{u=0}|-\pi .  \label{27n}
\end{equation}

As expected, the deflection angle in the weak limit approximation is found
to be the same result found by GBT. Namely, we have 
\begin{equation}
\hat{\alpha}\simeq \frac{\pi b_{0}^{2}}{4\,b^{2}}-\frac{3\pi Q^{2}}{4\,b^{2}}%
+\mathcal{O}(Q^{2},b_{0}^{2}).  \label{28n}
\end{equation}

\section{Deflection of Light by a Non-gravitational Wormhole}

Recently, using a non-linear electromagnetic theory in the framework of
Born-Infeld electromagnetism, it was shown that a specific field
configuration in flat metric can be viewed as a spherically symmetric
wormhole. In particular, the following effective spacetime metric was found
by \cite{Baldovin} as follows 
\begin{equation}
\mathrm{d}s^{2}=-\mathrm{d}t^{2}+\frac{\mathrm{d}r^{2}}{1-\frac{b(r)}{r}}+r^{2}\left( \mathrm{d}\theta ^{2}+\sin
^{2}\theta \,\mathrm{d}\varphi ^{2}\right) ,  \label{29ngwm}
\end{equation}%
where the shape function $b(r)$ is given by 
\begin{equation}
b(r)=\frac{2{r_{th}}^{4}-r^{4}+r^{2}\sqrt{r^{4}-{r_{th}}^{4}}}{r^{3}+r\sqrt{%
r^{4}-{r_{th}}^{4}}},  \label{30n}
\end{equation}%
by which $r_{th}$ is the throat location. The corresponding optical metric
of Eq. \eqref{29ngwm} can be found to be 
\begin{equation}
\mathrm{d}t^{2}=\frac{\mathrm{d}r^{2}}{1-\frac{2{r_{th}}^{4}-r^{4}+r^{2}%
\sqrt{r^{4}-{r_{th}}^{4}}}{r^{4}+r^{2}\sqrt{r^{4}-{r_{th}}^{4}}}}+r^{2}%
\mathrm{d}\varphi ^{2}.  \label{31opngwm}
\end{equation}

Using Eq. \eqref{7curvature} the Gaussian optical curvature for wormhole %
\eqref{31opngwm} becomes

\begin{equation}
\mathcal{K}=-\frac{2\left( r^{2}{r_{th}}^{4}-r^{6}+r^{4}\sqrt{r^{4}-{r_{th}}^{4}}+{%
r_{th}}^{4}\sqrt{r^{4}-{r_{th}}^{4}}\right) }{r^{4}\left( r^{2}+\sqrt{r^{4}-{%
r_{th}}^{4}}\right) \sqrt{r^{4}-{r_{th}}^{4}}}.  \label{32Gcw}
\end{equation}

In the following section, we shall use result \eqref{32Gcw} within the
concept of GBT to find the deflection angle. Next, we can approximate this
solution as follows 
\begin{equation}
\mathcal{K}\simeq -\frac{3{r_{th}}^{4}}{2\,r^{6}}-\frac{5{r_{th}}^{8}}{8r^{10}}
\label{33n}
\end{equation}

Substituting Eq. \eqref{33n} into Eq. \eqref{17n}, we get

\begin{equation}
\hat{\alpha}=-\int\limits_{0}^{\pi }\int\limits_{\frac{b}{\sin \varphi }%
}^{\infty }\left( -\frac{3{r_{th}}^{4}}{2r^{6}}-\frac{5{r_{th}}^{8}}{8r^{10}}%
\right) \sqrt{\det h_{ab}}\,\mathrm{d}r\mathrm{d}\varphi .
\label{34intx2}
\end{equation}

One can easily solve this integral in the leading order terms as 
\begin{equation}
\hat{\alpha}\simeq \frac{9\pi {r_{th}}^{4}}{64\,b^{4}}-\frac{175\pi {r_{th}}%
^{8}}{8192b^{8}}.  \label{35n}
\end{equation}

In the case of magnetic wormhole $\alpha =0$ and $\beta \neq 0$, we have $%
r_{th}=\sqrt{2\beta |/b_{0}}$. On the other hand, in the case of electric
wormhole ($\beta =0$ and $\alpha \neq 0$): $r_{th}=2\sqrt{2|\alpha |/b_{0}}$%
. Thus, we remark that in leading order terms the deflection angle is
affected by the non-linear electromagnetic fields. For magnetic field case: 
\begin{equation}
\hat{\alpha}\simeq \frac{9\pi \beta ^{2}}{16\,b^{4}\,b_{0}^{2}}+\mathcal{O}%
(\beta ^{4},b^{8}),  \label{36n}
\end{equation}%
and for electric field case: 
\begin{equation}
\hat{\alpha}\simeq \frac{9\pi \alpha ^{2}}{b^{4}\,b_{0}^{2}}+\mathcal{O}%
(\alpha ^{4},b^{8}).  \label{37n}
\end{equation}

\subsection{Geodesics of a Non-gravitational Wormhole}

For the geodesics analysis, one can write down the Lagrangian of metric %
\eqref{29ngwm} as follows 
\begin{eqnarray}\label{geo1}\notag
2\,\mathcal{L}&=&-\dot{t}^2(s)+\frac{\dot{r}(s)^2}{1-\frac{2{r_{th}}^4-r(s)^4+r(s)^2\sqrt{r(s)^4-{r_{th}}^4}}{r(s)^4+r(s)^2 \sqrt{r(s)^4-{r_{th}}^4}}}\\
&+&r(s)^2\dot{\theta}^2(s)+r(s)^2\sin^2\theta \dot{\varphi}^2(s).
\end{eqnarray}

Following Sec. (II-A), we first derive the two constants of
motion of the geodesics in the non-gravitational wormhole:

\begin{eqnarray}
\mathcal{G} &=&2r^{2}(s)\dot{\varphi}(s),  \label{39n} \\
\mathcal{E} &=&2\dot{t}(s).  \label{40n}
\end{eqnarray}

In sequel, without loss of generality, we normalize the affine parameter
along the light rays by taking $\mathcal{E}=1$ \cite{geodesicsBoyer} and
approximate the distance to the closest point by an impact parameter i.e., $%
u_{max}=1/r_{min}=1/b$ (recall that we consider only the leading order terms 
\cite{Iorio}). In this case, one can choose the second constant of motion as 
$\mathcal{G}=b$. Finally using the metric condition ($\epsilon =0$) together
with Eqs. (38-40), one can derive the following Lagrange equation: 
\begin{equation}
\left( \frac{\mathrm{d}u}{\mathrm{d}\varphi }\right) ^{2}\frac{1}{\left( 1-%
\frac{2r_{th}^{4}}{\Xi }+\frac{1}{\Xi u^{4}}-\frac{1-r_{th}^{4}u^{4}}{\Xi
u^{4}}\right) u^{4}}+\frac{1}{u^{2}}=0,  \label{41n}
\end{equation}

where 
\begin{equation}
\Xi =\frac{1+\sqrt{1-r_{th}^{4}u^{4}}}{u^{4}}.  \label{42n}
\end{equation}

After some manipulation, Eq. \eqref{41n} admits the following differential
equation

\begin{equation}
\frac{\mathrm{d}\varphi }{\mathrm{d}u}=\pm \frac{b\,b_{0}(\sqrt{\Delta }+1)}{%
\sqrt{2\left( 4\beta ^{2}u^{4}-b_{0}^{2}\right) \left( b^{2}u^{2}-1\right) (%
\sqrt{\Delta }+1)}},  \label{43n}
\end{equation}

where 
\begin{equation}
\Delta =\frac{b_{0}^{2}-4\beta ^{2}u^{4}}{b_{0}^{2}}.  \label{44n}
\end{equation}

After evaluating the integral, the leading order terms lead to the following
result for the magnetic field case: 
\begin{equation}
\hat{\alpha}\simeq \frac{9\pi \beta ^{2}}{16\,b^{4}\,b_{0}^{2}}.  \label{45n}
\end{equation}

If we compute the deflection angle of the non-gravitational wormhole
possessing pure electric field, we get 
\begin{equation}
\hat{\alpha}\simeq \frac{9\pi \alpha ^{2}}{b^{4}\,b_{0}^{2}}.  \label{46n}
\end{equation}

One can easily observe that Eqs. \eqref{45n} and \eqref{46n} are nothing but
the deflection angles obtained in the weak limit approximation of GBT.

\section{Deflection angle of Wormhole with Scalar Field}

Here we take the spacetime geometry to be that of a wormhole with scalar
field with metric \cite{sung} 
\begin{equation}
\mathrm{d}s^{2}=-\mathrm{d}t^{2}+\left( 1-\frac{b(r)}{r}+\frac{\alpha }{r^{2}}\right)
^{-1}\mathrm{d}r^{2}+r^{2}\left( \mathrm{d}\theta ^{2}+\sin ^{2}\theta \mathrm{d}\varphi ^{2}\right) .
\label{47wsfm}
\end{equation}

We will restrict our attention to the case $b(r)=b_{0}^{2}/r$, which
converts the metric \eqref{47wsfm} to the following form

\begin{equation}
\mathrm{d}s^{2}=-\mathrm{d}t^{2}+\left( 1-\frac{b_{0}^{2}}{r^{2}}+\frac{\alpha }{r^{2}}\right)
^{-1}\mathrm{d}r^{2}+r^{2}\left( \mathrm{d}\theta ^{2}+\sin ^{2}\theta \mathrm{d}\varphi ^{2}\right) .
\label{48}
\end{equation}

From here on in, one can immediately derive the optical metric as follows

\begin{equation}
\mathrm{d}t^{2}=\frac{\mathrm{d}r^{2}}{\left( 1-\frac{b_{0}^{2}}{r^{2}}+%
\frac{\alpha }{r^{2}}\right) }+r^{2}\mathrm{d}\varphi ^{2}.  \label{49}
\end{equation}

The Gaussian optical curvature \eqref{7curvature} of metric \eqref{49} is
then found to be 
\begin{equation}
\mathcal{K}={\frac{-b_0^2+\alpha }{{r}^{4}}}.  \label{50}
\end{equation}

If we substitute this equation into the deflection angle formula \eqref{17n}%
, we obtain 
\begin{equation}
\hat{\alpha}=-\int\limits_{0}^{\pi }\int\limits_{\frac{b}{\sin \varphi }%
}^{\infty }\left({\frac{-b_0^2+\alpha }{{r}^{4}}} \right) 
\sqrt{\det h_{ab}}\,\mathrm{d}r\mathrm{d}\varphi .  \label{51}
\end{equation}

After approximating $u \approx r$ and evaluating the integration the leading order terms yield

\begin{equation}
\hat{\alpha}\simeq \frac{\pi \,b_0^2}{4{b}^{2}}-{\frac{\pi
\,\alpha }{4{b}^{2}}}.  \label{52}
\end{equation}

Hence we conclude the effect of the scalar field is to bend the light outward the wormhole, in fact similar to the effect of electric charge.

\subsection{Geodesics of Wormhole with Scalar Field}

With the following Lagrangian, one can begin to perform the geodesics
analysis of the wormhole metric with scalar field \eqref{48}. 
\begin{equation}
2\mathcal{L}=-\dot{t}^{2}+\frac{\dot{r}^{2}}{1-\frac{b_{0}^{2}}{r^{2}}+\frac{%
\alpha }{r^{2}}}+r^{2}\dot{\theta}^{2}+r^{2}\sin ^{2}\theta \dot{\varphi}%
^{2}.  \label{53}
\end{equation}

Considering the planar photons ($\theta =\pi /2$) and taking cognizance of
the two constants of motion $l$ and $\gamma $: 
\begin{equation}
\mathcal{G}=2r^{2}(s)\dot{\varphi}(s),\text{ \ and \ }\mathcal{E}=2\dot{t}%
(s),  \label{54}
\end{equation}

we can normalize the affine parameter along the light rays by taking $%
\mathcal{E=}1$ (metric condition) \cite{geodesicsBoyer}. Again, we consider
the nearest approach distance with the impact parameter i.e., $u_{max}\simeq
1/b$ and use the second constant of motion $\mathcal{G}=b$ to focus on the
leading order terms \cite{Iorio}. In sequel, we use the metric condition
with Eqs. \eqref{53} and \eqref{54} and employ the Lagrange equations. Thus,
we obtain the following equation of motion for $u(\varphi )$: 
\begin{equation}
\left( \frac{\mathrm{d}u}{\mathrm{d}\varphi }\right) ^{2}\frac{1}{(\alpha
u^{2}-b_{0}^{2}u^{2}+1)u^{4}}-\frac{1}{u^{4}b^{2}}+\frac{1}{u^{2}}=0.
\label{55}
\end{equation}

From here, it follows that

\begin{equation}
\frac{\mathrm{d}\varphi }{\mathrm{d}u}=\pm \frac{b}{\sqrt{%
b^{2}(b_{0}^{2}u^{4}-\alpha u^{4}-u^{2})-b_{0}^{2}u^{2}+\alpha u^{2}+1}}.
\label{56}
\end{equation}

Solution of the differential equation \eqref{56} can be expressed as follows 
\cite{geodesicsBoyer} 
\begin{equation}
\Delta \varphi =\pi +\hat{\alpha},  \label{57}
\end{equation}%
where $\hat{\alpha}$ is the deflection angle to be calculated. The deflection angle results in 
\begin{equation}
\hat{\alpha}\simeq \frac{\pi \,b_0^2}{4{b}^{2}}-{\frac{\pi
\,\alpha }{4{b}^{2}}}.  \label{59}
\end{equation}

\section{Quantum Corrected Wormhole}

\subsection{Einstein-Rosen type wormhole}

As it is well-known, we can introduce a quantum velocity field $u_{\alpha }$
with a wave function of a quantum fluid as \cite{Ali1} 
\begin{equation}
\psi (x^{\alpha })=\mathcal{R}e^{i\,S(x^{\alpha })},  \label{60}
\end{equation}%
where $\psi (x^{\alpha })$ is a normalizable wave function, $\mathcal{R}%
(x^{\alpha })$ and $S(x^{\alpha })$ are some real continuous functions
associated with the four velocity field $u_{\alpha }=\frac{\hbar }{m}%
\,\partial _{\alpha }S$, in which $(\alpha =0,1,2,3)$. In a recent work, due
to implications of quantum corrections, Ali and Khalil \cite{Ali1} presented
an interesting model for a black hole solution known as modified
Schwarzschild black hole, when they replaced quantal (Bohmian) geodesics
with classical geodesics like the metric: 
\begin{equation}
\mathrm{d}s^{2}=-\left( 1-\frac{2M}{r}+\frac{\hbar \eta }{r^{2}}\right) \mathrm{d}t^{2}+\frac{%
\mathrm{d}r^{2}}{\left( 1-\frac{2M}{r}+\frac{\hbar \eta }{r^{2}}\right) }+r^2\mathrm{d}\Omega
_{2}^{2},  \label{61}
\end{equation}%
where $\eta $ is a dimensionless constant and $\hbar $ has a dimension of
(length)$^{2}$ in geometric units. When we consider $M=0$ and introducing a
new coordinate transformation $r^{2}=u^{2}+\mathcal{\epsilon }^{2}$, where $%
\mathcal{\epsilon }^{2}=-\hbar \eta ,$ the above metric reduces to the
Einstein-Rosen wormhole type solution: 
\begin{equation}
\mathrm{d}s^{2}=-\frac{u^{2}}{u^{2}-\hbar \eta }\mathrm{d}t^{2}+\mathrm{d}u^{2}+\left( u^{2}-\hbar \eta
\right) (\mathrm{d}\theta ^{2}+\sin ^{2}\theta \mathrm{d}\varphi ^{2}).  \label{62}
\end{equation}

This metric represents a massless quantum version of ERB wormhole with the
radius of the throat: $R_{thro.}=\sqrt{\hbar \eta }$. It is worth noting
that ERB wormhole is non-singular in the interval $u\in (-\infty ,\infty )$.
However this wormhole is a non-traversable wormhole. Its optical metric can
be written as 
\begin{equation}
\mathrm{d}t^{2}=\frac{u^{2}-\hbar \eta }{u^{2}}\mathrm{d}u^{2}+\frac{(u^{2}-\hbar \eta )^{2}}{%
u^{2}}\mathrm{d}\varphi ^{2}.  \label{63}
\end{equation}

The corresponding Gaussian optical curvature reads 
\begin{equation}
\mathcal{K}\simeq \frac{\hbar }{2r^{3}}+\frac{3\hbar ^{2}}{2r^{4}}.  \label{64}
\end{equation}

After some manipulation, one can compute the deflection angle \eqref{17n} as 
\begin{equation}
\hat{\alpha}\simeq -\frac{\hbar }{b}-\frac{3\hbar ^{2}\pi }{8b^{2}}.
\label{65}
\end{equation}

One can infer from the above result that since the deflection of light is
negative, \textit{it indicates that light rays always bend outward the
wormhole due to the quantum correction effects}. Of course, for the resulting expression to  the deflection angle we need to take the absolute value of the above equation
\begin{equation}
|\hat{\alpha}|\simeq \frac{\hbar }{b}+\frac{3\hbar ^{2}\pi }{8b^{2}}.
\end{equation}

It is well known that the ER wormholes are not traversable wormholes, hence a static traversable Morris-Thorne wormholes have been introduced \cite{Morris}.

\subsection{Morris-Thorne traversable wormhole}

The spacetime ansatz for seeking static spherically symmetric traversable
(without event horizon) wormhole spacetime can be written in the
Schwarzschild coordinates as \cite{Morris} 
\begin{equation}
\mathrm{d}s^{2}=-e^{2\Phi (r)}\mathrm{d}t^{2}+\frac{\mathrm{d}r^{2}}{1-\frac{b(r)}{r}}+r^{2}\left(
\mathrm{d}\theta ^{2}+\sin ^{2}\theta \mathrm{d}\varphi ^{2}\right) ,  \label{66}
\end{equation}%
where ($t,r,$ $\theta $, $\varphi $) are the standard Schwarzschild
coordinates, with $\Phi (r)$ and $b(r)$ are the redshift and shape
functions, respectively. Since we seek a wormhole solution, the redshift
function $\Phi (r)$ should be finite everywhere, in order to avoid the
presence of an event horizon. Another essential characteristics of a
wormhole geometry is about the shape function $b(r)$, which should satisfy
the condition of $b(r_{0})=r_{0}$, in which $r_{0}$ is the radius of the
wormhole throat where the two spacetimes (upper/lower spacetimes) are
joined. Consequently, the shape function must satisfy the flaring-out
condition: 
\begin{equation}
\frac{b(r)-rb^{\prime }(r)}{b^{2}(r)}>0,  \label{67}
\end{equation}%
in which $b^{\prime }(r)=\frac{db}{dr}<1$ must hold at the throat of the
wormhole. In addition to this, the shape function also satisfy the
asymptotically flat limit i.e., $b(r)/r\rightarrow 0$ as $r\rightarrow
\infty $.

From now on, we focus on the effect of quantum corrections. Einstein
equations remain the same with the effect of quantum corrections in the
energy-momentum tensor: 
\begin{equation}
G_{\mu \nu }=R_{\mu \nu }-\frac{1}{2}g_{\mu \nu }R=8\pi T_{\mu \nu }^{eff},
\label{68}
\end{equation}%
where $G_{\mu \nu }$ and $T_{\mu \nu }^{eff}$ are the Einstein tensor and
effective energy-momentum tensor, respectively. The total-derivative of the
source term in the field equations is written as 
\begin{equation}
T_{\mu \nu }^{eff}=T_{\mu \nu }+T_{\mu \nu }^{corr.}.  \label{69}
\end{equation}%
For an anisotropic fluid $T_{\mu \nu }$, the energy-momentum tensor is
defined as follows 
\begin{equation}
{{T^{\mu }}_{\nu }}=\left( -\rho ,\mathcal{P}_{r},\mathcal{P}_{\theta },%
\mathcal{P}_{\varphi }\right) ,  \label{70}
\end{equation}%
where $\rho $ is the energy density with $\mathcal{P}_{r}$, $\mathcal{P}%
_{\theta }$ and $\mathcal{P}_{\varphi }$ are non-zero components of the
diagonal terms, as measured in the orthogonal direction, respectively. The
second term of Eq. \eqref{69} has some modifications when the quantum
corrections are taken into consideration: 
\begin{equation}
{{T^{\mu }}_{\nu }}^{(corr.)}=\left( -\rho
^{(corr.)},P_{r}^{(corr.)},P_{\theta }^{(corr.)},P_{\varphi
}^{(corr.)}\right).  \label{71}
\end{equation}%

Using the definition of the 4-momentum given as follows
\begin{equation}
p_\alpha=\hbar \,\partial_\alpha S,
\end{equation}
using the wave function solution one can show that the geodesic motion is modified due to the relativistic quantum potential
given by
\begin{equation}
V_Q=\hbar^2 \frac{\Box \mathcal{R}}{\mathcal{R}}.
\end{equation}

In Ref. \cite{Ali1} it was shown that for a general and static spherically
symmetric spacetime metric in Schwarzschild coordinates, the stress-energy tensor has the following non-zero components
\begin{equation}
\rho ^{(corr.)}=P_{r}^{(corr.)}=P_{\theta }^{(corr.)}=P_{\varphi }^{(corr.)}=%
\frac{\hbar \eta }{8\pi r^{4}},  \label{72}
\end{equation}%
in which $\eta $ is dimensionless constant. It just happened that the effect of this quantum potential to be similar to the effect of the electric field described by a similar energy-momentum tensor components simply by letting $Q^{2}\rightarrow \hbar
\eta $. On the other hand, the only non-zero components form the Einstein tensor $G_{\mu \nu }$ of the metric %
\eqref{66} are 
\begin{eqnarray}
G_{t}^{t} &=&-\frac{b^{\prime }(r)}{r^{2}},  \notag \\
G_{r}^{r} &=&-\frac{b(r)}{r^{3}}+2\left( 1-\frac{b(r)}{r}\right) \frac{\Phi
^{\prime }}{r},  \notag \\
G_{\theta }^{{\theta }} &=&\left( 1-\frac{b(r)}{r}\right) \Big[\Phi ^{\prime
\prime }+(\Phi ^{\prime })^{2}-\frac{b^{\prime }r-b}{2r(r-b)}\Phi ^{\prime }
\notag \\
&-&\frac{b^{\prime }r-b}{2r^{2}(r-b)}\Phi ^{\prime }+\frac{\Phi ^{\prime }}{r%
}\Big],  \notag \\
G_{\varphi }^{{\varphi }} &=&G_{\theta }^{{\theta }}.  \label{73n}
\end{eqnarray}%
Using \eqref{68} and the Morris-Thorne metric \eqref{66}, we obtain the
following set of field equations 
\begin{eqnarray}
\rho (r) &=&\frac{1}{8\pi r^{2}}\left[ b^{\prime }(r)-\frac{\hbar \eta }{%
r^{2}}\right] ,  \notag \\
\mathcal{P}_{r}(r) &=&\frac{1}{8\pi }\left[ 2\left( 1-\frac{b(r)}{r}\right) 
\frac{\Phi ^{\prime }}{r}-\frac{b(r)}{r^{3}}-\frac{\hbar \eta }{r^{4}}\right]
,  \notag \\
\mathcal{P}(r) &=&\frac{1}{8\pi }\left( 1-\frac{b(r)}{r}\right) \Big[\Phi
^{\prime \prime }+(\Phi ^{\prime })^{2}-\frac{b^{\prime }r-b}{2r(r-b)}\Phi
^{\prime }  \notag \\
&-&\frac{b^{\prime }r-b}{2r^{2}(r-b)}\Phi ^{\prime }+\frac{\Phi ^{\prime }}{r%
}-\frac{\hbar \eta }{r^{4}}\Big].  \label{74n}
\end{eqnarray}%
where $\mathcal{P}=\mathcal{P}_{\theta }=\mathcal{P}_{\varphi }$ is the
pressure measured in the tangential directions. Meanwhile, one can observe
that the charged wormhole solution simply by letting $Q^{2}\rightarrow \hbar
\eta $ \cite{sung}. Moreover, the geometry encoded in the Einstein equations due to the quantum effect can be obtained by shifting the shape function in the following way 
\begin{equation}
b\rightarrow b_{eff}=b-\frac{\hbar \eta }{r}.  \label{75n}
\end{equation}%
With more preciously, the quantum effect on the wormhole reveals when
shifting the shape function $b$ into $(b-{\hbar \eta }/{r})$, after we make use of the following choose 
\begin{equation}
\Phi(r) \to \Upsilon(r) = \frac{1}{2}\ln\left(1+\frac{\hbar \eta}{r^2}\right).
\end{equation}

We can see this from Einstein's field equations written in terms of the effective shape function 
\begin{eqnarray}
\rho (r) &=&\frac{b_{eff}^{\prime }}{8\pi r^{2}},  \notag \\
\mathcal{P}_{r}(r) &=&\frac{1}{8\pi }\left[ 2\left( 1-\frac{b_{eff}(r)}{r}%
\right) \frac{\Upsilon ^{\prime }}{r}-\frac{b_{eff}(r)}{r^{3}}\right] , 
\notag \\
\mathcal{P}(r) &=&\frac{1}{8\pi }\left( 1-\frac{b_{eff}(r)}{r}\right) \Big[%
\Upsilon ^{\prime \prime }+(\Upsilon ^{\prime })^{2}-\frac{b_{eff}^{\prime
}r-b_{eff}}{2r(r-b_{eff})}\Upsilon ^{\prime }  \notag \\
&-&\frac{b_{eff}^{\prime }r-b_{eff}}{2r^{2}(r-b_{eff})}\Upsilon ^{\prime }+%
\frac{\Upsilon ^{\prime }}{r}\Big],  \label{76}
\end{eqnarray}%
where 
\begin{eqnarray}\notag
\Upsilon ^{\prime } &=&-\frac{\hbar \eta }{r^{3}}\left( 1+\frac{\hbar \eta }{%
r^{2}}\right) ^{-1},  \notag \\
\Upsilon ^{\prime \prime } &=&\frac{\hbar \eta }{r^{4}}\left( 1+\frac{\hbar
\eta }{r^{2}}\right) ^{-1}\left[ 3-\frac{2\hbar \eta }{r^{2}}\left( 1+\frac{%
\hbar \eta }{r^{2}}\right) ^{-1}\right] .  \label{77}
\end{eqnarray}

Our obtained solution refurbishes the work of charged wormhole solution in
which it was assumed a probable charged metric in a combination of MT type
static spherically symmetric wormhole and Reissner-Nordstr$\ddot{\text{o}}$%
m(RN) spacetime. We now demonstrate that our result, due to the quantum
effects, is similar to that procedure. The key point is that we can recast
the spacetime metric as 
\begin{equation}
\mathrm{d}s^{2}=-\left( 1+\frac{\hbar \eta }{r^{2}}\right) \mathrm{d}t^{2}+\frac{dr^{2}}{1-%
\frac{b(r)}{r}+\frac{\hbar \eta }{r^{2}}}+r^2\mathrm{d}\Omega _{2}^{2},  \label{78}
\end{equation}%
which is a combination of Morris-Thorne type wormhole with quantum
corrections. 


\subsubsection{Gravitational lensing}

Let us now continue our study to investigate the deflection of light in the quantum corrected spacetime geometry given by Eq.\eqref{78}. Choosing $b(r)=b_{0}^{2}/r$, metric \eqref{78} reduces to 
\begin{equation}
\mathrm{d}s^{2}=-\left( 1+\frac{\hbar \eta }{r^{2}}\right) \mathrm{d}t^{2}+\frac{\mathrm{d}r^{2}}{1-%
\frac{b_{0}^{2}}{r^{2}}+\frac{\hbar \eta }{r^{2}}}+r^2\mathrm{d}\Omega _{2}^{2}.
\label{79}
\end{equation}

The quantum corrected optical Gaussian curvature therefore yields 
\begin{equation}
\mathcal{K}\simeq \frac{3\hbar \eta -b_{0}^{2}}{r^{4}}-\frac{4b_{0}^{2}\hbar \eta }{%
r^{6}}.  \label{80}
\end{equation}

Thus, the asymptotic form of the geodesic curvature becomes 
\begin{equation}
\lim_{R\rightarrow \infty }\kappa (C_{R})\rightarrow \frac{1}{R},  \label{81}
\end{equation}

which leads the following deflection angle: 
\begin{equation}
\hat{\alpha}\simeq \frac{\pi b_{0}^{2}}{4\,b^{2}}-\frac{3\pi \hbar \eta }{%
4\,b^{2}}+\frac{3\,\pi \hbar \eta \,b_{0}^{2}}{8b^{4}}.  \label{82}
\end{equation}

\subsubsection{Energy Conditions}

In this section, we argue the issue of energy conditions and make some
regional plots to check the validity of all energy conditions. WEC is
defined by $T_{\mu \nu }U^{\mu }U^{\nu }\geq 0$ i.e., $\rho \geq 0$ and $%
\rho (r)+\mathcal{P}_{r}(r)\geq 0$, where $T_{\mu \nu }$ is the energy
momentum tensor and $U^{\mu }$ denotes the timelike vector. This means that
local energy density is positive and it gives rise to the continuity of NEC,
which is defined by $T_{\mu \nu }k^{\mu }k^{\nu }\geq 0$ i.e., $\rho (r)+%
\mathcal{P}_{r}(r)\geq 0$, where $k^{\mu }$ is a null vector. In this
regard, we have the following energy condition at the throat region: 
\begin{equation}
\rho (r_{0})=\frac{b_{eff}^{\prime }(r_{0})}{8\pi r_{0}^{2}}.  \label{83}
\end{equation}%
Now, using the field equations \mref{73n,74n,75n,76,77}, one finds the
following relations 
\begin{equation}
\rho (r)+\mathcal{P}_{r}(r)=\frac{1}{8\pi }\left[ 2\left( 1-\frac{b_{eff}}{r}%
\right) \frac{\Upsilon ^{\prime }}{r}+\frac{rb_{eff}^{\prime }-b_{eff}}{r^{3}%
}\right] ,  \label{84}
\end{equation}%
and 
\begin{eqnarray}
\rho (r)+\mathcal{P}(r) &=&\frac{1}{8\pi }\Big[\left( 1-\frac{b_{eff}(r)}{r}%
\right) \Big(\Upsilon ^{\prime \prime }+(\Upsilon ^{\prime })^{2}  \notag \\
&-&\frac{b_{eff}^{\prime }r-b_{eff}}{2r(r-b_{eff})}\Upsilon ^{\prime }-\frac{%
b_{eff}^{\prime }r-b_{eff}}{2r^{2}(r-b_{eff})}\Upsilon ^{\prime }  \notag \\
&+&\frac{b_{eff}^{\prime }}{r^{2}}+\frac{\Upsilon ^{\prime }}{r}\Big)\Big].
\label{85}
\end{eqnarray}

\begin{figure}[h!]
\center
\includegraphics[width=0.54\textwidth]{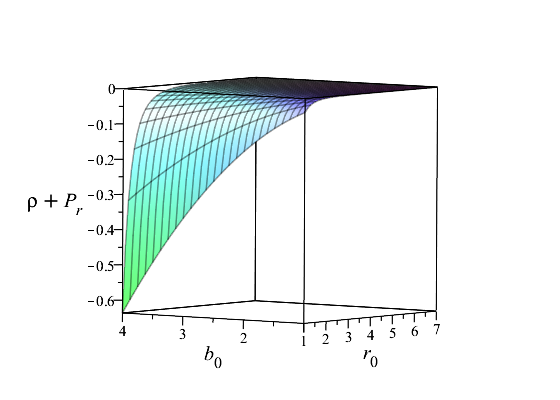}
\caption{{\protect\small \textit{ The figure shows the behavior of the NEC i.e.,
$\rho+\mathcal{P}_r$ as a function of $r_0$ and $b_0$, for chosen $\hbar=1$ and $\eta=1$. We see that NEC is violated. }}}
\label{f1}
\end{figure}

For an observation one can see from Eq.\eqref{75n}, the solutions are asymptotically flat, i.e.,
$\frac{b_{eff}(r)}{r} \rightarrow 0$ as r $\rightarrow \infty$. With the above 
solution one also verifies that the form function obeys 
$b'_{eff}(r_0) = \left(\hbar \eta-b_0^2 \right)/r^2 < 1$, for flaring out condition at the throat.
Taking into account the condition we should impose the restriction $\hbar \eta < 2r_0^2$.
As when we consider the wormhole throat, Eq.\eqref{84} reduces to
 \begin{eqnarray}
\left(\rho+\mathcal{P}_r\right)|_{r=r_0} = \frac{1}{8\pi r_0^4}\Big[-\frac{2(\hbar \eta)^2}{r_0^2+\hbar \eta}
+2(\hbar \eta-r_0^2)\Big].
 \end{eqnarray}
Taking into account the condition $\hbar \eta < 2r_0^2$, one verifies that matter configuration violates
NEC at the throat i.e. $\left(\rho+\mathcal{P}_r\right)|_{r=r_0}< 0$. In Fig. (\ref{f1}), we show the behavior of the null energy
condition.

\subsubsection{Volume Integral Quantifier}
We now  consider the ``volume integral quantifier," which provides information
about measure of the “amount of exotic matter required” for wormhole maintenance
and is related only to $\rho$  and $\mathcal{P}_r$, not to the transverse components,  with the following definite integrals $I_V =\int\left(\rho(r)+\mathcal{P}_r(r)\right)\mathrm{d}V$, where $\mathrm{d}V=r^2\sin\theta \mathrm{d}r \mathrm{d}\theta \mathrm{d}\phi$.
Taking into account Eq.\eqref{84} and performing the integration by parts, which is
\begin{eqnarray}
\oint [\rho+\mathcal{P}_r]~\mathrm{d}V  &=& \left[(r-b_{eff})\ln\Big(\frac{1+\hbar \eta/r^2}{1-b_{eff}/r}\Big)  \right]_{r_0}^{\infty}\\\notag
&-&\int_{r_0}^{\infty}\left[(r-b'_{eff})\ln\Big(\frac{1+\hbar \eta/r^2}{1-b_{eff}/r}\Big)  \right]\mathrm{d}r.
\end{eqnarray}
For the construction of wormhole the boundary term at $r_0$ vanishes, and due to 
asymptotic behavior the boundary term at infinity also vanishes. Thus the above expression reduces to
\begin{equation}
\oint [\rho+\mathcal{P}_r]~\mathrm{d}V = -\int_{r_0}^{\infty}\left[(r-b'_{eff})\ln\Big(\frac{1+\hbar \eta/r^2}{1-b_{eff}/r}\Big)  \right]\mathrm{d}r.
\end{equation}
The value of this volume-integral provides information about the ``total amount"
of ANEC violating matter in the spacetime, and we are going to solve the integral for 
our particular choice of shape function $b(r)=b^2_0/r$. Here, we suppose that  wormhole extends form $r_0$ to a radius situated at $`a'$ and then 
we get the very simple result
\begin{eqnarray}\label{whfe1}
I_V=\frac{b_0^2}{\sqrt{\hbar \eta}}\left[\text{arctan}\left(\frac{r_0}{\sqrt{\hbar \eta}} \right)-
\text{arctan}\left(\frac{a}{\sqrt{\hbar \eta}} \right)\right].
\end{eqnarray}
For an interesting observation when $a \rightarrow r_0$ then $\int{(\rho+\mathcal{P}_r)} \rightarrow 0$, and thus one may interpret that wormhole can be contracted for with arbitrarily small quantities of ANEC violating matter.

\section{Conclusions}
In this paper, we have pointed out in detail the effect of electric charge, magnetic charge, scalar fields, and quantum effects on the deflection angle in the various wormhole geometries. To this end,  by considering the optical geometries of the wormholes, we have employed the GWM in the context of GBT. Then, the deflection angles are computed via the surface integrals of the associated Gaussian optical curvatures on a domain outside the light ray, which reveals the effect of gravitational lensing as a global effect of the spacetime.  \\

For the wormholes having electric/scalar fields, the deflection angles are found out to be proportional to the throat of the wormhole that bands the light rays towards itself. On the other hand, it is seen that electric/scalar fields play an important role to band the light rays outward the wormhole center. In the case of a non-linear electromagnetic theory in which the wormhole geometry is constructed purely by non-linear magnetic and electric fields, we have shown that deflection angle is proportional to the electromagnetic fields. \\

We have also constructed the Einstein-Rosen type wormhole solutions as well as the Morris-Thorne wormhole, which are both supported by the quantum effects coming from the Bohmian quantum mechanics. In particular, it has been shown that the exotic matter concentrated on the throat of the wormhole violates the null energy conditions. It is worth noting that the quantum effects on the wormhole geometry are correlated with the shape function: shifting of the shape function is in complete analogy with the charge wormhole solution. \\

Finally, we want to emphasize that although our results show consistency in the first-order terms in the deflection angles, the harmonized results of the methods disappear with the consideration of the second-order
correction terms. This is simply due to the straight line approximation used in the integration domain. In near future, 
we plan to commit a study that will show fit in the results of the second-order correction terms. The latter aim is going to be possible when we manage to produce a suitable equation for the light ray or to modify the integration domain associated to the optical geometry.  \\

\begin{acknowledgments}
This work was supported by the Chilean FONDECYT Grant No. 3170035 (A\"{O}). A%
\"{O} is grateful to the Waterloo University, Department of Physics and
Astronomy and Perimeter Institute for Theoretical Physics for hosting him as
a research visitor where part of this work was done.

\end{acknowledgments}

\end{document}